\documentclass[12pt,a4paper]{article}

\usepackage[T1]{fontenc}
\usepackage{lmodern}
\usepackage{amsmath}
\usepackage{amsfonts}
\usepackage{amssymb}
\usepackage{mathrsfs}
\usepackage{physics}
\usepackage{graphicx}
\usepackage{caption}
\usepackage{subcaption}
\usepackage{jheppub}
\usepackage{enumitem}
\usepackage{framed,float}
\usepackage{booktabs}
\usepackage{array}
\usepackage{soul}
\usepackage{multirow}
\usepackage{comment}
\usepackage{braket}
\usepackage{todonotes}
\usepackage{youngtab}
\usepackage{tikz}
\usetikzlibrary{calc,arrows,decorations.markings}

\usepackage{color}

\preprint{}
\title{\boldmath Decoupling Limit of Quiver Theories and the Angular Spectra of Extreme C-metrics}

\author[a]{Peng Yang}
\author[a,b,c,1]{and Kilar Zhang\note{Corresponding Author.}}
\affiliation[a]{Department of Physics and Institute for Quantum Science and Technology, Shanghai University, 99 Shangda Road, Shanghai 200444, China}
\affiliation[b]{Shanghai Key Lab for Astrophysics, Shanghai Normal University, 100 Guilin Road, Shanghai 200234, China}
\affiliation[c]{Shanghai Key Laboratory of High Temperature Superconductors, Shanghai 200444, China}
\emailAdd{yangpeng522@shu.edu.cn}
\emailAdd{kilar@shu.edu.cn}

\abstract{
We investigate the angular eigenvalue problem of the extreme charged C-metric. Starting from a constrained five-point Fuchsian description of the non-extremal angular equation, we obtain its extreme limit by decoupling a hypermultiplet in the dual four-dimensional $\mathcal{N}=2$, $\mathrm{SU(2)}\times \mathrm{SU(2)}$ linear quiver gauge theory. We derive the parameter dictionary and renormalized Matone relations, and connect the physical limit $Q\to M$ to the quiver decoupling by an explicit coordinate transformation. Regular endpoint conditions give an algebraic quantization condition. Eigenvalues obtained from the truncated instanton prepotential are tested against independent angular spectra near extremality and at larger acceleration. The axisymmetric sector recovers the known exact spectrum.
}

\begin{document}

\maketitle
\flushbottom

\setcounter{footnote}{0}

\section{Introduction}
\label{sec:introduction}

The study of black hole perturbation theory provides crucial insights into the stability and holographic properties of gravitational systems. Among various exact solutions, the charged C-metric \cite{Kinnersley-Walker} describes accelerating black holes and exhibits intriguing phase structures in the context of the AdS/CFT correspondence \cite{Anabalon:2018ydc,Boido:2022iye,Cassani:2021dwa,Arenas-Henriquez:2022www,Arenas-Henriquez:2023hur}. Calculating the exact quasinormal modes for this background is technically challenging because the acceleration parameter breaks standard spherical symmetry. As a result, the angular differential equation no longer admits standard spherical harmonics as its solutions \cite{Griffiths:2005qp,griffiths2009,Griffiths:2006tk}. The angular equation admits a standard Heun representation \cite{Chen:2024cmassless}. In the extreme black hole limit ($Q\to M$), the inner and outer horizons coalesce. For nonzero azimuthal number the associated two angular singularities fuse into a rank-1 irregular singular point, giving a confluent Heun equation; the axisymmetric sector further reduces to a hypergeometric equation \cite{Kofron:2015separability}.

Recently, exact analytical methods based on the Alday-Gaiotto-Tachikawa (AGT) correspondence \cite{Alday:2009aq} and the Nekrasov-Shatashvili (NS) limit \cite{Nekrasov:2009rc} have been extensively applied to black hole perturbation theory. The gauge-theory input builds on Seiberg--Witten theory \cite{Seiberg:1994rs} and instanton counting \cite{Nekrasov:2002qd,Nekrasov:2003rj}. In this framework, the perturbation equations are treated as quantum Seiberg-Witten curves, and their connection coefficients are exactly computed via the instanton partition functions of four-dimensional $\mathcal{N}=2$ supersymmetric gauge theories (see, for example, \cite{Aminov:2020yma,Bonelli:2021uvf,Casals:2021ugr,Bianchi:2021mft, Aminov:2023jve,Aminov:2024mul} and recent developments in \cite{Hatsuda:2020sbn, Consoli:2022eey, Fioravanti:2021dce, Grassi:2022zuk, Imaizumi:2021cxf, Amado:2021erf, Dodelson:2020lal, Novaes:2018fry, Lisovyy:2021bkm, Wang:2026kue, Lei:2023mqx, Yang:2026xrz, Ge:2025yqk}). Surface-defect constructions relate Heun equations directly to class S theories \cite{Jeong:2018qpc}, while convergence properties of Nekrasov functions have also been studied rigorously \cite{Arnaudo:2022ivo}. For Fuchsian equations with more than four regular singularities, exact connection formulae have been systematically formulated using linear quiver gauge theories. In particular, recent work in \cite{Arnaudo:2025kof} detailed the procedure for computing connection coefficients for generic five-puncture topologies, which complements the intersecting surface defect constructions used to solve 5-point Fuchsian systems and degenerate Knizhnik-Zamolodchikov equations without quantization ambiguities \cite{Jeong:2020uxz,Jeong:2021rll}. This algebraic structure is deeply intertwined with the framework of quivers and spectral networks in the presence of black holes \cite{Arnaudo:2025kof}. Concurrently, the analytical structure of irregular Liouville correlators, arising when regular singularities coalesce, has been developed in \cite{Lisovyy:2018mnj, Bonelli:2022ten} to solve classical confluent Heun equations. Related Heun-to-confluent-Heun limits have also been used to obtain exact low-temperature black-hole Green's functions \cite{Arnaudo:2024sen}. 

The angular equation of the extreme C-metric provides a concrete setting for the degeneration of regular conformal blocks into irregular ones through horizon fusion. Its constrained five-point description allows us to follow this process within the $\mathrm{SU(2)}\times\mathrm{SU(2)}$ linear quiver.

In this paper, we develop the angular quiver description outlined in the earlier study of C-metric quasinormal modes \cite{Lei:2023mqx}, including its extreme charged limit. We first establish the regular parameter dictionary and connection problem. We then construct the confluent equation and Nekrasov prepotential by decoupling a hypermultiplet, derive the renormalized accessory parameters, and connect this construction to the physical limit $Q\to M$ by an explicit change of coordinates. The connection coefficients yield the angular quantization condition, from which we evaluate the nonaxisymmetric spectrum and recover the axisymmetric result of \cite{Kofron:2015separability}. Independent angular calculations test the instanton approximations and the continuity of the spectrum towards extremality.

The paper is organized as follows. Section \ref{sec:generic_c_metric} reviews the angular equation for the generic charged C-metric and evaluates its eigenvalues using the regular quiver gauge theory. Section \ref{sec:confluent_quiver} develops the theoretical framework for the confluent quiver theory, detailing the decoupling limit of the Nekrasov prepotential and verifying the analytical framework through the computation of the local connection coefficients. Section \ref{sec:extreme_c_metric} applies this confluent framework to the extreme charged C-metric to evaluate the angular spectra. Section \ref{sec:conclusion} concludes the paper and outlines several directions for future research. Appendix \ref{app:combinatorics} gives the confluent equations, prepotentials and Matone relations \cite{Matone:1995rx} for arbitrary linear quivers, together with the coordinate and normalization relations used in the angular calculations.

\section{Angular Spectra of the Generic Charged C-metric}
\label{sec:generic_c_metric}

The complete evaluation of quasinormal modes requires solving both the radial and angular equations. The earlier work \cite{Lei:2023mqx} developed the QNM framework and radial analysis, and outlined a quiver treatment of the angular equation alongside numerical spectra and small-acceleration expansions. Here we develop that angular description and its confluent limit. Consequently, our primary focus in this paper is strictly on the angular eigenvalue problem. Before addressing the confluent limit, we first formulate the regular setup by studying the angular equation of the generic (non-extreme) charged C-metric. This establishes the parameter dictionary and the exact quantization condition from the regular $\mathrm{SU(2)} \times \mathrm{SU(2)}$ quiver gauge theory, which serves as the basis for the singularity fusion in the subsequent sections.

\subsection{Angular Equation and Singularity Structure}

We consider a neutral massless scalar on the charged C-metric with vanishing cosmological constant \cite{Lei:2023mqx}. In the coordinates used here,
\begin{equation}
\begin{aligned}
 ds^2&=\Omega^{-2}\left[-f(r)d\tau^2+\frac{dr^2}{f(r)}
 +\frac{r^2d\theta^2}{P(\theta)}+r^2P(\theta)\sin^2\theta\,d\varphi^2\right],\\
 f(r)&=\left(1-\frac{2M}{r}+\frac{Q^2}{r^2}\right)(1-\alpha^2r^2),
 \qquad\Omega=1-\alpha r\cos\theta.
\end{aligned}
\end{equation}
Away from the conical axis $R[g]=0$, so $\Box_g\Phi=0$ is equivalent to
$(\Box_g-R[g]/6)\Phi=0$. Writing $g=\Omega^{-2}\widetilde g$ and
$\Phi=\Omega\widetilde\Phi$ gives
$(\Box_{\widetilde g}-R[\widetilde g]/6)\widetilde\Phi=0$.
The ansatz $\widetilde\Phi=e^{-i\omega\tau+im\varphi}R(r)\chi(\theta)/r$
separates this equation. We choose the azimuthal period
$\Delta\varphi=2\pi/P(\pi)$, which removes the conical defect at $\theta=\pi$.
Since the angular equation depends on $Q^2$, we take $0<Q\le M$ and assume $0<\alpha r_+<1$, where
$r_+=M+\sqrt{M^2-Q^2}$, so $P(\theta)>0$ on the physical angular interval.
The zero-acceleration limit is taken continuously.

 The angular equation is given by
\begin{equation}\label{eq:angular_original}
    P(\theta)\sin\theta \frac{\mathrm{d}}{\mathrm{d}\theta} \left( P(\theta)\sin\theta \frac{\mathrm{d}\chi}{\mathrm{d}\theta} \right) + (V_\theta(\theta) - m^2)\chi = 0 \,,
\end{equation}
where $m = m_0 P(\pi)$ with the azimuthal quantum number $m_0 \in \mathbb{Z}$, and the metric function $P(\theta)$ is defined as
\begin{equation}
    P(\theta) = 1 - 2\alpha M \cos\theta + \alpha^2 Q^2 \cos^2\theta \,.
\end{equation}
The effective potential $V_\theta(\theta)$ reads
\begin{equation}
    V_\theta(\theta) = P(\theta) \left( \lambda \sin^2\theta - \frac{P(\theta)\sin^2\theta}{3} + \frac{\sin\theta\cos\theta P'(\theta)}{2} + \frac{\sin^2\theta P''(\theta)}{6} \right) \,,
\end{equation}
where $\lambda$ is the separation constant (the angular eigenvalue) to be determined. 

Introducing the algebraic variable $x = \cos\theta$, we can rewrite \eqref{eq:angular_original} in the standard Schrödinger-like form:
\begin{equation}\label{eq:generic_schrodinger}
    \frac{\mathrm{d}^2\psi}{\mathrm{d}x^2} + Q_x(x)\psi = 0 \,,
\end{equation}
where $ \chi(x) = (P(x)(1-x^2))^{-1/2} \psi(x) $.

With $H(x)=(1-x^2)P(x)$, the original equation and its normal-form potential are
\begin{equation}\label{eq:normal_potential}
 (H\chi')'+\left(\lambda+\frac{H''}{6}-\frac{m^2}{H}\right)\chi=0,
 \qquad
 Q_x=\frac{\lambda}{H}-\frac{H''}{3H}
      +\frac{H'^2/4-m^2}{H^2}.
\end{equation}
In particular, $\alpha\to0$ gives $\lambda\to\ell(\ell+1)+1/3$ in this separation-constant convention.

In the generic case ($Q < M$), the polynomial $P(x)$ has two distinct real roots
\begin{equation}
    y_\pm = \frac{M \pm \sqrt{M^2-Q^2}}{\alpha Q^2} \,.
\end{equation}
The four finite points $x=\pm1,y_\pm$ are regular singular points. For quartic $H$, the $x^{-2}$ and $x^{-3}$ coefficients in \eqref{eq:normal_potential} cancel, so $Q_x=O(x^{-4})$. Under $w=1/x$, the normal-form wavefunction is $\widetilde\psi(w)=w\psi(1/w)$ up to a constant, and the potential is $w^{-4}Q_x(1/w)$, which is analytic at $w=0$. Thus $x=\infty$ is ordinary, and the physical equation has four genuine regular singularities. 

We retain the five marked points as a constrained specialization of the general five-point BPZ system. Via the AGT correspondence this embeds \eqref{eq:generic_schrodinger} into the $\Omega$-deformed 4d $\mathcal N=2$, $\mathrm{SU(2)}\times\mathrm{SU(2)}$ linear quiver framework \cite{Arnaudo:2025kof,Bonelli:2022ten}; the extra ordinary point is not an additional physical singularity.

We place the marked points in the conformal coordinate $z$ at $\{0,t,1,q,\infty\}$ and use the weak-coupling sewing parameters $t$ and $p=1/q$. The chamber $0<|t|<1<|q|<\infty$ admits the nested-channel expansion used below.

We apply a global Möbius transformation to map the physical domain $x \in [-1, 1]$ to the interval $[0, t]$ while relocating the remaining singularities. A convenient choice of mapping is
\begin{equation}\label{eq:mobius_regular}
    z(x) = \frac{t(x+1)}{2} \,, \quad \text{with} \quad t = \frac{2}{y_- + 1} \,.
\end{equation}
Under this transformation, the marked points $\{ -1, 1, y_-, y_+, \infty \}$ are mapped to $\{ 0, t, 1, q, \infty \}$ respectively, where $q = \frac{y_++1}{y_-+1}$. 
We adopt the transformation from \cite{Lei:2023mqx} with a simple modification, ensuring the expansion parameter satisfies $|1/q| < 1$.

In the $z$-coordinate, the transformed potential $Q(z) = Q_x(x(z)) (\mathrm{d}x/\mathrm{d}z)^2$ takes the canonical Fuchsian form (also referred to as the Extended Heun Equation (EHE) by \cite{Yang:2026xrz}) 
    \begin{equation}
\begin{aligned}
 Q(z)&=\sum_{i=0,1,t,q}\frac{\frac14-a_i^2}{(z-z_i)^2}
 +\frac{\frac14-a_\infty^2-\sum_{i=0,1,t,q}(\frac14-a_i^2)}{z(z-1)}\\
 &\quad+\frac{(t-1)u_t}{z(z-1)(z-t)}
 +\frac{(q-1)u_q}{z(z-1)(z-q)}\,.
\end{aligned}
    \end{equation}
The local indicial equations determine the squared gauge theory mass parameters $a_i^2$.
For the real physical domain $y_+>y_->1$, we choose their
nonnegative representatives:
\begin{equation}\label{eq:mass_dictionary}
\begin{aligned}
    a_0 &=
    \frac{|m|y_+y_-}{2(1+y_+)(1+y_-)}
    \,, &
    a_t &=
    \frac{|m|y_+y_-}{2(y_+-1)(y_--1)}
    \,,\\
    a_1 &=
    \frac{|m|y_+y_-}{(y_-^2-1)(y_+-y_-)}
    \,, &
    a_q &=
    \frac{|m|y_+y_-}{(y_+^2-1)(y_+-y_-)}
    \,,\qquad
    a_\infty=\frac12.
\end{aligned}
\end{equation}
The labels follow the puncture correspondence
$x=-1,1,y_-,y_+\mapsto z=0,t,1,q$, respectively.
The accessory parameters $u_t$ and $u_q$, which encode the eigenvalue $\lambda$, are determined by the residues of the potential:
\begin{align}
     u_t &= -\frac{5 - 3y_+ - 3y_- + y_+y_-}{6(1-y_+)(1-y_-)} - \frac{y_+y_-}{(1-y_-)(1-y_+)}\lambda \nonumber\\ &\quad+ \frac{y_+^2 y_-^2 \big[ 5 + y_- (y_+ - 3) - 3y_+ \big]}{2(1-y_+)^3 (1-y_-)^3}m^2 \,, \label{eq:ut} \\
     u_q &= -\frac{1 - 3y_+^2 + 2y_+y_-}{6(1-y_+)(y_+-y_-)} + \frac{y_+y_-}{(1-y_+)(y_+-y_-)}\lambda \nonumber\\ &\quad+ \frac{2y_+^2 y_-^2 (1 - 3y_+^2 + 2y_+y_-)}{(1-y_+)^3 (1+y_+)^2 (y_+-y_-)^3}m^2 \,. \label{eq:uq}
\end{align}
These variables are related to the Coulomb branch parameters $b_1$ and $b_2$ of the gauge theory via the inverted Matone relations, which can be solved perturbatively in $t$ and $1/q$. The underlying prepotential--modulus relation was established in \cite{Matone:1995rx}, related to the nonperturbative renormalization group in \cite{Bonelli:1996rg}, and extended to include gravitational couplings in \cite{Flume:2004rp}.

Writing $\Delta_i=1/4-a_i^2$, the physical dictionary obeys
\begin{equation}\label{eq:ordinary_constraint}
 \Delta_\infty=0,\qquad
 2(t\Delta_t+\Delta_1+q\Delta_q)-\sum_{i=0,t,1,q}\Delta_i
 +(t-1)u_t+(q-1)u_q=0.
\end{equation}
These are the vanishing $z^{-2}$ and $z^{-3}$ coefficients at the ordinary point.
In the conformal-block description, the ordinary point is a vacuum insertion. At finite $\epsilon_2=\varepsilon$, it is represented by $a_\infty=(1+\varepsilon)/2$. Removing the vacuum state fixes the adjacent internal module to the external module at $q$, giving $b_2^2=a_q^2$ in the NS limit. We use the branch $b_2=a_q$ continuous from weak coupling.

\subsection{Connection Problem and Quantization Condition}

To determine the physical bound states, we must impose appropriate boundary conditions on the angular wavefunction. The physical angular domain $\theta \in [0, \pi]$ maps to $x \in [1, -1]$. Under the Möbius transformation, the poles at $x=-1$ and $x=1$ are relocated to $z=0$ and $z=t$, respectively. Physical requirements dictate that the wavefunction must remain regular at both poles.

In terms of the wavefunction $\psi(z)$, the Frobenius solutions near the boundaries behave as
\begin{equation}\label{eq:bc_frob}
\begin{aligned}
    &\psi(z) = \psi_{0,+}(z) \sim z^{\frac{1}{2}+a_0} &\text{for}\ z \sim 0\,, \\
    &\psi(z) = \psi_{t,+}(z) \sim (z-t)^{\frac{1}{2}+a_t} &\text{for}\ z \sim t\,.
\end{aligned}
\end{equation}
Here, we have selected the positive signs $+a_0$ and $+a_t$ to denote the branches corresponding to the regular (non-divergent) physical modes at the respective poles, with the branch convention in \eqref{eq:mass_dictionary},
$a_0,a_t>0$ for $m\neq0$.
Since the transformation from $\chi$ to $\psi$ contributes
a square-root factor at each physical endpoint, the two local
behaviours of the angular wavefunction are
$\chi\sim(1+x)^{\pm a_0}$ near $x=-1$ and
$\chi\sim(1-x)^{\pm a_t}$ near $x=1$.
Regularity therefore selects $\psi_{0,+}$ and $\psi_{t,+}$.
Choosing the opposite sign for an endpoint mass would exchange
the corresponding Frobenius labels, without changing the
physical boundary condition.
For $m_0=0$, the exponents coincide and regularity instead
requires the absence of the logarithmic solution, as discussed
in Appendix~\ref{app:l0_proof}.

For every physical $m_0$, the dictionary also gives $2a_0=|m_0|\in\mathbb Z$.
The nonresonant connection formula is therefore understood by analytic
continuation of the complete coefficient. For $m_0\ne0$ the coefficient of the
leading singular endpoint behaviour has a finite continuation proportional to
$\Gamma(2a_0)$; the coincident-exponent case $m_0=0$ requires the logarithmic
prescription described below.

The exact connection formula between the solutions expanded around $z=0$ and $z=t$ is derived from the fusion properties of the conformal blocks. Following the approach in \cite{Arnaudo:2025kof}, the connection relation reads

\begin{equation}\label{eq:connection_formula}
\begin{aligned}
 \psi_{t,\theta}(z)&=\sum_{\theta'=\pm}C_{0,\theta'}^{t,\theta}\psi_{0,\theta'}(z),\\
 C_{0,\theta'}^{t,\theta}
 &=e^{i\pi(1/2+\theta a_t)}t^{\theta a_t-\theta'a_0}
 \frac{\Gamma(1+2\theta a_t)\Gamma(-2\theta'a_0)}
 {\displaystyle\prod_{\sigma=\pm}\Gamma(1/2+\theta a_t-\theta'a_0+\sigma b_1)}\\
 &\quad\times\exp\left[\frac\theta2\partial_{a_t}F_{\rm q}(p,t)
                  -\frac{\theta'}2\partial_{a_0}F_{\rm q}(p,t)\right].
\end{aligned}
\end{equation}
Here $F_{\rm q}=F_{\rm inst}+F_{U(1)}$ is the normalized, tree-subtracted
instanton prepotential. The one-loop contribution is written explicitly as
Gamma functions. Mass derivatives are taken at fixed internal Coulomb parameters
before imposing the Matone relations; the combinatorial convention is given in
Appendix~\ref{app:combinatorics}.

Regularity at $z=0$ requires the coefficient of the singular branch $\psi_{0,-}$ in the continuation of $\psi_{t,+}$ to vanish. Setting $\theta = +$ for the regular mode at $z=t$ and demanding the coefficient of $\theta' = -$ to be zero, we obtain the condition

\begin{equation}
 \frac{\Gamma(1+2a_t)\Gamma(2a_0)}
 {\displaystyle\prod_{\sigma=\pm}\Gamma(1/2+a_t+a_0+\sigma b_1)}
 \exp\left[\frac12(\partial_{a_t}+\partial_{a_0})F_{\rm q}\right]=0.
\end{equation}
At generic internal momenta in the weak-coupling domain, the exponential is
finite and nonzero. Its analytic continuation to the physical branch is taken
only after combining the complete expressions. The zeros then arise from the
reciprocal one-loop Gamma factors
\begin{equation}
 \mathcal G=\prod_{\sigma=\pm}
 \frac1{\Gamma(1/2+\sigma b_1+a_t+a_0)}.
\end{equation}

Thus, the connection coefficient vanishes when the argument of the Gamma function reaches a non-positive integer. Selecting the proper Weyl branch (by symmetry $b_1 \to -b_1$), this yields the algebraic quantization condition
\begin{equation}\label{eq:qc_regular_final}
    b_1(\lambda) = \frac{1}{2} + a_t + a_0 + n \,, \quad n \in \mathbb{Z}_{\ge 0} \,,
\end{equation}
where $n = \ell - |m_0|$ corresponds to the number of nodes of the angular wavefunction. 

We apply the quantization condition \eqref{eq:qc_regular_final} to evaluate the angular eigenvalues for the generic C-metric. 
The parameter $b_1$ is a function of the physical parameters and the eigenvalue $\lambda$. Following \cite{Arnaudo:2025kof}, it can be obtained by inverting the Matone relations
\begin{equation}
\begin{aligned}
    &u_t = -\frac{1}{4} - b_1^2 + a_0^2 + a_t^2 + t\frac{\partial F_{\rm q}\left(\frac{1}{q},t\right)}{\partial t} \,, \\
    &u_q = -\frac{1}{4} + b_2^2 - a_{\infty}^2 + a_q^2 + q\frac{\partial F_{\rm q}\left(\frac{1}{q},t\right)}{\partial q} \,.
\end{aligned}
\end{equation}

For $m_0=0$, the endpoint exponents coincide
($a_0=a_t=0$), so the nonresonant connection coefficients
cannot be evaluated by direct substitution.
Instead, regularity requires the absence of the logarithmic
Frobenius solution.
The corresponding coefficient is obtained from the regulated
combination
$D_{\log}=\lim_{a_t\to0}\lim_{a_0\to0}a_0(C_+-C_-)$,
as explained in Appendix~\ref{app:l0_proof}.
In the normalized conformal-block convention,
$F_{\rm q}=F_{\rm inst}+F_{U(1)}$ is even in $a_0$ and $a_t$
at generic internal momenta, so its first derivatives with
respect to these parameters vanish at the origin.
The regulated connection condition therefore gives
$b_1=\ell+1/2$, with resonant internal momenta understood
by analytic continuation of the complete expressions.
Unlike the extreme case considered in the appendix,
$F_{\rm q}$ generally remains nonzero here, and its coupling
derivatives must be retained in the Matone relations.

On the physical vacuum locus, we evaluate the normalized prepotential
using $T=t(1-p)/(1-pt)$, with $p=1/q$:

\begin{equation}\label{eq:reduced_heun_F}
 \mathcal F_4(T)=\lim_{\varepsilon\to0}\varepsilon\log Z_4(T)
 -2(a_1+\tfrac12)(a_t+\tfrac12)\log(1-T),
\end{equation}
where $Z_4$ is the Young-diagram sum with hypermultiplet masses
$a_t+a_0,a_t-a_0,a_1+a_q,a_1-a_q$. The coordinate and normalization relations to the five-point prepotential
are given in Appendix~\ref{app:physical_reductions}.

For the spectrum we fix
$b=1/2+a_0+a_t+\ell-|m_0|$ and evaluate
\begin{equation}
 u_H=-\frac14-b^2+a_0^2+a_t^2+T\partial_T\mathcal F_4(T).
\end{equation}
The transformed accessory parameter is related to $u_t$ by
\begin{equation}\label{eq:heun_accessory}
 u_H=(1-pt)u_t-2pt\left(\frac14-a_t^2\right).
\end{equation}
Together with \eqref{eq:ut}, this determines $\lambda$. Table~\ref{tab:generic_cmetric} displays orders $k=3,6,10$ in $T$, with independent angular Galerkin benchmarks. Here $k$ is the instanton degree in $T$. The quantized momentum is fixed before truncating the prepotential, and the calculation details are given in Appendix~\ref{app:numerics}.

\begin{table}[htbp]
\centering\footnotesize
\setlength{\tabcolsep}{4pt}\renewcommand{\arraystretch}{1.18}
\begin{tabular}{cccccc}\toprule
$(m_0,\ell)$ & $k=3$ & $k=6$ & $k=10$ & Numerical & $|\delta\lambda_{10}|$\\
\midrule
\multicolumn{6}{c}{$\alpha M=0.05,\quad Q/M=0.3$} \\
\midrule
$(0,0)$ & 0.3318158785 & 0.3317380765 & 0.3317375480 & 0.3317375474 & $5.61\times10^{-10}$ \\
$(0,1)$ & 2.3240103832 & 2.3245939856 & 2.3245964693 & 2.3245964712 & $1.90\times10^{-9}$ \\
$(0,2)$ & 6.3080696887 & 6.3099584343 & 6.3099668247 & 6.3099668314 & $6.71\times10^{-9}$ \\
$(0,3)$ & 12.2842135975 & 12.2880629216 & 12.2880801895 & 12.2880802034 & $1.39\times10^{-8}$ \\
$(1,1)$ & 2.6452750934 & 2.6461039492 & 2.6461077470 & 2.6461077502 & $3.13\times10^{-9}$ \\
$(1,2)$ & 6.8507230068 & 6.8529370461 & 6.8529471267 & 6.8529471349 & $8.24\times10^{-9}$ \\
$(2,2)$ & 7.3726123128 & 7.3754469589 & 7.3754605288 & 7.3754605404 & $1.16\times10^{-8}$ \\
$(2,3)$ & 13.7912191259 & 13.7961612369 & 13.7961843495 & 13.7961843688 & $1.94\times10^{-8}$ \\
\midrule
\multicolumn{6}{c}{$\alpha M=0.10,\quad Q/M=0.8$} \\
\midrule
$(0,0)$ & 0.3289627433 & 0.3287772680 & 0.3287748811 & 0.3287748752 & $5.93\times10^{-9}$ \\
$(0,1)$ & 2.3039163477 & 2.3052941318 & 2.3053052846 & 2.3053053046 & $2.00\times10^{-8}$ \\
$(0,2)$ & 6.2533121834 & 6.2577743877 & 6.2578120883 & 6.2578121590 & $7.07\times10^{-8}$ \\
$(0,3)$ & 12.1774911658 & 12.1865868143 & 12.1866644155 & 12.1866645624 & $1.47\times10^{-7}$ \\
$(1,1)$ & 2.9994736685 & 3.0021665614 & 3.0021907365 & 3.0021907844 & $4.79\times10^{-8}$ \\
$(1,2)$ & 7.4381822467 & 7.4443464480 & 7.4444005261 & 7.4444006311 & $1.05\times10^{-7}$ \\
$(2,2)$ & 8.5480170651 & 8.5574647573 & 8.5575525882 & 8.5575527673 & $1.79\times10^{-7}$ \\
$(2,3)$ & 15.4493769103 & 15.4641843501 & 15.4643182344 & 15.4643185012 & $2.67\times10^{-7}$ \\
\bottomrule\end{tabular}
\caption{Generic angular eigenvalues at the indicated instanton orders $k$ in $T$. The error is the absolute difference between the
$k=10$ approximation and the independent angular eigenvalue. }
\label{tab:generic_cmetric}
\end{table}
The approximations approach the independent angular eigenvalues as the instanton order increases. 

\section{Decoupling Limit of the Quiver Gauge Theory}
\label{sec:confluent_quiver}

We now construct a confluent extension of the regular \(\mathrm{SU(2)}\times\mathrm{SU(2)}\) quiver framework. We start from the general five-point canonical equation, with the singular points placed at \(0,t,1,q,\infty\), and take a double-scaling limit in which the points at \(q\) and infinity coalesce. The parameters in this construction are treated independently of the C-metric dictionary introduced in the preceding section. This choice places the resulting irregular singularity at infinity and provides a canonical equation for the subsequent application to the extreme angular problem.

The decoupling limit is defined by
\[
a_q=\frac{\eta-m_{\mathrm{SW}}}{2},\qquad
a_\infty=\frac{-\eta-m_{\mathrm{SW}}}{2},\qquad
q=\frac{\eta}{L},\qquad \eta\to\infty,
\]
with \(t,a_0,a_t,a_1,L,m_{\mathrm{SW}}\) held fixed. Writing
\[
K=\sum_{i=0,t,1}\left(\frac14-a_i^2\right),
\]
we keep \(u_t\) finite and choose the accessory parameter \(u_q\) such that
\[
u_q=-\eta m_{\mathrm{SW}}-K-u_0,
\]
where \(u_0\) remains finite. The terms involving the colliding singularities can then be reorganized as
\[
\frac{u_0}{z(z-1)}
+\frac{u_q}{z(z-q)}
+\frac{\frac14-a_q^2}{(z-q)^2}.
\]
At fixed \(z\), away from the remaining finite poles, their limit is
\[
\frac{u_0}{z(z-1)}
+\frac{m_{\mathrm{SW}}L}{z}
-\frac{L^2}{4}.
\]
Together with the unchanged finite-pole terms, this gives the confluent equation below:
{\footnotesize
\begin{equation}\label{eq:irregular_ODE}
    \left[ \partial_z^2 + \frac{\frac{1}{4}-a_0^2}{z^2} + \frac{\frac{1}{4}-a_t^2}{(z-t)^2} + \frac{\frac{1}{4}-a_1^2}{(z-1)^2} + \frac{u_0}{z(z-1)} + \frac{(t-1)u_t}{z(z-1)(z-t)} - \frac{L^2}{4} + \frac{m_{SW} L}{z} \right] \psi(z) = 0 \,.
\end{equation}
}
The accessory parameter $u_0$ fixes the finite coefficient of $1/[z(z-1)]$ surviving the collision, and is obtained by regulating the diverging bare parameters:
\begin{equation}\label{eq:u0_limit}
    u_0 = \lim_{\eta \to \infty} \left[ a_q^2 - a_\infty^2 - \left( \frac{1}{4}-a_0^2 + \frac{1}{4}-a_1^2 + \frac{1}{4}-a_t^2 \right) - u_q \right] \,.
\end{equation}

Via the AGT correspondence, this CFT collision limit translates directly to the decoupling limit of a fundamental hypermultiplet in the right $\mathrm{SU(2)}$ node of the $\mathrm{SU(2)}\times \mathrm{SU(2)}$ linear quiver, reducing the effective flavor count of that node from $N_f=4$ to $N_f=3$ \cite{Gaiotto:2009ma,Marshakov:2009agt}. The tree-subtracted NS prepotential $F^{\text{NS}}=F_{\rm q}+F_{\text{1-loop}}$ degenerates accordingly.

For the instanton partition function, the structural modification occurs exclusively at the second gauge node. The two hypermultiplet factors depending on $a_q$ and $a_\infty$ collapse into a single term with mass $-m_{SW}$. The degenerate instanton sum evaluates as:
\begin{equation}
\begin{aligned}
    Z_{inst} &= \sum_{\vec{Y}, \vec{W}} t^{|\vec{Y}|} L^{|\vec{W}|} z_{vec}(\vec{b}_1, \vec{Y}) z_{vec}(\vec{b}_2, \vec{W}) z_{bif}(\vec{b}_1, \vec{Y}, \vec{b}_2, \vec{W}; -a_1) \\
    &\quad \times \left( \prod_{\sigma=\pm} z_{hyp}(\vec{b}_1, \vec{Y}, a_t + \sigma a_0) \right) z_{hyp}(\vec{b}_2, \vec{W}, -m_{SW}) \,,
\end{aligned}
\end{equation}
where $\vec{b}_1, \vec{b}_2$ denote the Coulomb branch parameters. The explicit combinatorial definitions ($z_{vec}, z_{bif}, z_{hyp}$) are documented in Appendix \ref{app:combinatorics}.

The $U(1)$ normalization factor also degenerates. As the expansion parameter $p = L/\eta \to 0$, the $p$-dependent logarithmic series truncates exactly to a linear function of $L$:
\begin{equation}
    F_{U(1)} = -2\left(a_1+\frac{1}{2}\right)\left(a_t+\frac{1}{2}\right)\ln(1-t) + \left(\frac{1}{2}-a_1\right)L + \left(a_t+\frac{1}{2}\right)t L \,.
\end{equation}
The first-node and bifundamental masses and Coulomb parameters are held fixed in this limit. Their one-loop terms therefore remain unchanged. The heavy right-node hypermultiplet contribution is absorbed into the finite-part prescription \eqref{eq:one_loop_finite_part}.

The first-node Matone relation retains its form because the pole at $z=t$ remains fixed in the collision. Substituting \eqref{eq:u0_limit} into the regular second-node relation cancels the heavy-mass terms and gives
\begin{equation}\label{eq:renormalized_matone}
\begin{cases}
 b_1^2=-\frac14+a_0^2+a_t^2-u_t+t\partial_tF_{\rm q}(t,L),\\[8pt]
 b_2^2=-\frac12+a_0^2+a_1^2+a_t^2-u_0+L\partial_LF_{\rm q}(t,L).
\end{cases}
\end{equation}
The one-loop terms are independent of the sewing parameters, so these derivatives act on $F_{\rm q}=F_{\rm inst}+F_{U(1)}$.

\subsection{Numerical Verification}
We first test the confluent connection formula between the regular singular points at $z=0$ and $z=t$, before applying it to the angular eigenvalue problem. Taking the decoupling limit of the regular connection formula, with the same local Frobenius normalization and branch conventions, preserves its Gamma-function structure, while replacing the prepotential and the internal parameter $b_1$ by their confluent counterparts. The connection coefficient $\mathcal{C}_{0,\theta_0}^{t,\theta_t}$ is thus given by:
\begin{equation}
\begin{aligned}
    \mathcal{C}_{0, \theta_0}^{t, \theta_t} &= \exp\left[ i\pi\left(\frac{1}{2} + \theta_t a_t\right) \right] t^{\theta_t a_t - \theta_0 a_0} \\
    &\quad \times \frac{\Gamma(1+2\theta_t a_t) \Gamma(-2\theta_0 a_0)}{\Gamma\left(\frac{1}{2} + \theta_t a_t - \theta_0 a_0 + b_1\right) \Gamma\left(\frac{1}{2} + \theta_t a_t - \theta_0 a_0 - b_1\right)} \\
    &\quad \times \exp\left[ \frac{\theta_t}{2}\partial_{a_t} F_{\rm q}\left(t, L\right) - \frac{\theta_0}{2}\partial_{a_0} F_{\rm q}\left(t, L\right) \right] \,.
\end{aligned}
\end{equation}

Here $F_{\rm q}=F_{\rm inst}+F_{U(1)}$ denotes the normalized confluent instanton prepotential, and its mass derivatives are taken at fixed internal Coulomb parameters before substituting the solutions of the Matone relations.

To test the confluent connection formula and the decoupling limit numerically, we fix the target parameters directly as
\begin{equation*}
    t=\frac{1}{100}\,,\quad a_0=\frac{97}{70}\,,\quad
    a_1=\frac{7}{141}\,,\quad a_t=\frac{10}{9}\,,\quad u_t=\frac{1}{50}\,,
\end{equation*}
\begin{equation*}
    L=\frac{7}{24000}\,,\qquad m_{\mathrm{SW}}=-\frac{313}{120}\,,\qquad
    u_0=\frac{6128537651}{2805606720}\,.
\end{equation*}
We compare the Nekrasov results with independent Wronskian calculations, following the numerical method of \cite{Arnaudo:2025kof}. At order $N$, we truncate $F_{\rm q}$ to total instanton degree $N$, solve the two truncated Matone equations directly on the branch connected to $N=0$, and evaluate its mass derivatives at fixed Coulomb parameters. Gamma functions and exponentials are not re-expanded. Table~\ref{tab:irregular_connection} compares $N=0,\ldots,6$ with independent Frobenius--Wronskian results. This target tests the generic confluent quiver before specialization to the C-metric.

For a separate check of the decoupling limit, the regular five-point parameters are reconstructed at each finite $\eta$ from the same fixed target:
\begin{equation}
\begin{aligned}
    q_\eta&=\frac{\eta}{L}\,, &
    a_{q,\eta}&=\frac{\eta-m_{\mathrm{SW}}}{2}\,,\\
    a_{\infty,\eta}&=\frac{-\eta-m_{\mathrm{SW}}}{2}\,, &
    u_{q,\eta}&=-\eta m_{\mathrm{SW}}-K-u_0\,,
\end{aligned}
\end{equation}
where $K=(1/4-a_0^2)+(1/4-a_1^2)+(1/4-a_t^2)$. Table~\ref{tab:regular_decoupling} compares the six-instanton, directly solved regular Nekrasov result for $\mathcal{C}_{0,+}^{t,-}$ with the EHE Wronskian result along this family. As $\eta$ increases, the two columns approach the independently computed six-instanton confluent result and the CEHE Wronskian result, respectively. Their remaining difference in the limit reflects the finite instanton truncation.

\begin{table}[htbp]
\centering\footnotesize
\renewcommand{\arraystretch}{1.25}
\begin{tabular}{ccc}\toprule
$N$ & $\mathcal C_{0,+}^{t,-}$ & $\mathcal C_{0,-}^{t,-}$\\\midrule
0 & $-142813.86214244-392377.86142145i$ & $0.03884950457+0.10673813655i$ \\
1 & $-143618.50213053-394588.59161954i$ & $0.04046222978+0.11116906265i$ \\
2 & $-143630.39572629-394621.26900530i$ & $0.04048003943+0.11121799428i$ \\
3 & $-143630.52440329-394621.62254245i$ & $0.04048024129+0.11121854887i$ \\
4 & $-143630.52628831-394621.62772151i$ & $0.04048024457+0.11121855788i$ \\
5 & $-143630.52633141-394621.62783994i$ & $0.04048024465+0.11121855811i$ \\
6 & $-143630.52633289-394621.62784399i$ & $0.04048024465+0.11121855812i$ \\
\midrule
Wronskian & $-143630.52633295-394621.62784417i$ & $0.04048024465+0.11121855812i$\\
\bottomrule\end{tabular}
\caption{Confluent connection coefficients obtained by directly
solving the Matone relations truncated at total degree $N$. The independent
Wronskian values use normalized Frobenius series at both finite endpoints.}
\label{tab:irregular_connection}
\end{table}

\begin{table}[htbp]
\centering\small
\renewcommand{\arraystretch}{1.5}
\begin{tabular}{rcc}\toprule
$\eta$ & Regular Nekrasov ($N=6$) & Regular Wronskian\\\midrule
$10$ & \shortstack{$-143630.5286827444$\\$-394621.6343001696\,i$} & \shortstack{$-143630.5286828091$\\$-394621.6343003474\,i$} \\[4pt]
$1000$ & \shortstack{$-143630.5263563867$\\$-394621.6279085545\,i$} & \shortstack{$-143630.5263564515$\\$-394621.6279087324\,i$} \\[4pt]
$100000$ & \shortstack{$-143630.5263331233$\\$-394621.6278446387\,i$} & \shortstack{$-143630.5263331880$\\$-394621.6278448166\,i$} \\[4pt]
$\infty$ & \shortstack{$-143630.5263328883$\\$-394621.6278439931\,i$} & \shortstack{$-143630.5263329530$\\$-394621.6278441710\,i$} \\[4pt]
\bottomrule\end{tabular}
\caption{Decoupling of $\mathcal C_{0,+}^{t,-}$ at a fixed
confluent target. The same direct Matone prescription and total instanton
degree six are used at every finite $\eta$ and at the confluent limit.}
\label{tab:regular_decoupling}
\end{table}

\section{Angular Spectra of the Extreme Charged C-metric}
\label{sec:extreme_c_metric}

Applying the confluent quiver framework developed in Section \ref{sec:confluent_quiver}, we resolve the angular eigenvalue problem for the extreme charged C-metric. 

\subsection{Singularity Fusion and Parameter Dictionary}

In the extreme limit $Q \to M$, the metric function $P(\theta)$ reduces to a perfect square. Using the algebraic coordinate $x = \cos\theta$, we have
\begin{equation}
    P(x) = (1 - \alpha M x)^2 \,.
\end{equation}
The two regular singularities $y_\pm$ are related to the radial horizons by $y_-=1/(\alpha r_+)$ and $y_+=1/(\alpha r_-)$. They coalesce into a single double root at $x = y_0 \equiv \frac{1}{\alpha M}$. Substituting this into the effective angular potential $V_\theta(x)$ and factoring out $(1-x^2)$, the potential becomes:
\begin{equation}
    V_\theta(x) = (1-\alpha M x)^2 (1-x^2) \left[ \lambda - \frac{1}{3} + \frac{\alpha^2 M^2}{3} + 2\alpha M x - 2\alpha^2 M^2 x^2 \right] \,.
\end{equation}
For $m_0\ne0$, the fusion point $x=y_0=1/(\alpha M)$ becomes an irregular singular point of Poincar\'e rank 1. The equation is a constrained specialization of the CEHE in the previous section. For $m_0=0$ the irregular terms vanish, and the equation further reduces to a hypergeometric equation, as shown in Appendix~\ref{app:l0_proof}.

To match the irregular ODE \eqref{eq:irregular_ODE} derived from the confluent quiver theory, we use a Möbius transformation that sends the physical endpoints to two regular singular points and the irregular singularity to $z=\infty$. The transformation is defined as:
\begin{equation}
    z(x) = \frac{x + 1}{x - y_0} \quad \implies \quad x(z) = \frac{y_0 z + 1}{z - 1} \,,
\end{equation}
where $y_0=1/(\alpha M)$ is the double root of the angular polynomial. 
Under this mapping, the physical boundaries representing the south and north poles ($x=-1$ and $x=1$) are projected to $z=0$ and $z=t = \frac{2}{1-y_0}$, respectively. The potential in the $z$-coordinate transforms as a quadratic differential $Q(z) = Q_x(x(z)) (\mathrm{d}x/\mathrm{d}z)^2$, with $\psi_z(z)=(dx/dz)^{-1/2}\psi_x(x(z))$. The Schwarzian derivative vanishes for this M\"obius map. 

The parameter dictionary mapping the regular and irregular singularities is obtained as:
\begin{equation}\label{eq:ext_masses}
    a_0^2 = \frac{m^2 y_0^4}{4(y_0+1)^4} \,, \quad a_t^2 = \frac{m^2 y_0^4}{4(y_0-1)^4} \,, \quad a_1^2 = \frac{1}{4} \,,
\end{equation}
For the numerical calculations, we take the nonnegative roots
for $a_0$ and $a_t$ and choose $a_1=+1/2$.
\begin{equation}\label{eq:ext_irregular}
    L = \frac{2|m| y_0^2}{(y_0-1)(y_0+1)^2} \,, \qquad m_{SW} = \frac{2|m| y_0^3}{(y_0^2-1)^2} \,.
\end{equation}
The accessory parameters $u_t$ and $u_0$ are given by:
\begin{equation}\label{eq:ext_u0_ut}
    u_t = u_0 = -\frac{1}{6} - \frac{y_0^2}{y_0^2-1} \lambda + \frac{y_0^4(y_0-3)}{2(y_0-1)^4(y_0+1)} m^2 \,.
\end{equation}

\paragraph{Relation to the regular C-metric.}
Let $z_R,t_R,q_R$ denote the regular coordinates of Section~\ref{sec:generic_c_metric}, and use a superscript $(R)$ for their mass and accessory parameters. Before taking $Q\to M$, introduce
\begin{equation}\label{eq:physical_collision_frame}
 \widehat z=\frac{z_R}{z_R-1}=\frac{x+1}{x-y_-},\qquad
 \widehat t=\frac{t_R}{t_R-1},\qquad
 \widehat q=\frac{q_R}{q_R-1}.
\end{equation}
The old marked points $0,t_R,1,q_R,\infty$ map to $0,\widehat t,\infty,\widehat q,1$. In particular, the ordinary point is now at $\widehat z=1$, with $\widehat a_1=1/2$, while the two roots $y_-$ and $y_+$ lie at infinity and $\widehat q$. For $m_0\ne0$, choose
\begin{equation}\label{eq:physical_heavy_mass}
\begin{aligned}
 \eta&=a_1^{(R)}+a_q^{(R)},&
 \widehat m_{\rm SW}&=a_1^{(R)}-a_q^{(R)},\\
 \widehat a_q&=a_q^{(R)}=\frac{\eta-\widehat m_{\rm SW}}2,&
 \widehat a_\infty&=-a_1^{(R)}=\frac{-\eta-\widehat m_{\rm SW}}2,\\
 \widehat L&=\frac{\eta}{\widehat q}.
\end{aligned}
\end{equation}
The normal-form transformation gives
\begin{equation}\label{eq:physical_accessory_bridge}
\begin{aligned}
 \widehat u_t&=(1-t_R)u_t^{(R)}-2t_R\Delta_t^{(R)},\\
 \widehat u_q&=(1-q_R)u_q^{(R)}-2q_R\Delta_q^{(R)},\\
 \widehat u_0&=-\eta\widehat m_{\rm SW}
              -\widehat K-\widehat u_q=\widehat u_t,
 \qquad \widehat K=\Delta_0^{(R)}+\Delta_t^{(R)}.
\end{aligned}
\end{equation}
The last equality follows from \eqref{eq:ordinary_constraint}. As $y_+\to y_-=y_0$, we have $\eta,\widehat q\to\infty$ and
\begin{equation}\label{eq:physical_collision_limits}
\begin{aligned}
 \widehat t&\longrightarrow\frac2{1-y_0},&
 \widehat L&\longrightarrow\frac{2|m|y_0^2}{(y_0-1)(y_0+1)^2},\\
 \widehat m_{\rm SW}&\longrightarrow\frac{2|m|y_0^3}{(y_0^2-1)^2},&
 \widehat u_t=\widehat u_0&\longrightarrow u,
\end{aligned}
\end{equation}
where $u$ is the expression in \eqref{eq:ext_u0_ut}. This recovers the extreme dictionary from the regular physical family in the collision frame of Section~\ref{sec:confluent_quiver}. In the axisymmetric sector the irregular terms vanish, giving the hypergeometric reduction discussed below.

The extreme dictionary satisfies $a_1^2=1/4$ and $u_t=u_0=u$,
so the marked point at $z=1$ is ordinary. Its vacuum insertion identifies
the adjacent internal modules; we choose $b_1=b_2=b$ on the weak-coupling
branch. On this locus, the normalized prepotential depends on $\kappa=tL$:
\begin{equation}\label{eq:reduced_prepotential}
 F_{\rm q}(t,L;b,b)\big|_{a_1=1/2}=\mathcal F_3(tL;b),
 \qquad
 \mathcal F_3(\kappa;b)=\left(a_t+\frac12\right)\kappa
             +\lim_{\varepsilon\to0}\varepsilon\log Z_3(\kappa),
\end{equation}
Here $Z_3$ is the single-node instanton sum with hypermultiplet masses
$a_t+a_0$, $a_t-a_0$ and $-m_{\rm SW}$.
Its definition and the coordinate and normalization relations are given in
Appendix~\ref{app:physical_reductions}.
Both renormalized Matone relations give
\begin{equation}\label{eq:reduced_matone}
 b^2=-\frac14+a_0^2+a_t^2-u
       +\kappa\partial_\kappa\mathcal F_3(\kappa;b).
\end{equation}

\subsection{Boundary Conditions and Numerical Results}

The determination of the angular eigenvalue relies on formulating a connection problem between the two physical poles $\theta=\pi$ ($z=0$) and $\theta=0$ ($z=t$). A crucial observation is that the irregular singularity is located at $x = y_0 = \frac{1}{\alpha M}$, which, due to $\alpha M < 1$, resides outside the physical domain $x \in [-1, 1]$. 

The decoupling limit of the normalized connection coefficient in Section~\ref{sec:confluent_quiver} preserves its Gamma-function structure, while its prepotential and internal momenta change. For $t<0$ we use $\log t=\log|t|+i\pi$ and continue the local powers on the same branch; the resulting nonzero phase does not affect the zeros. Regularity at both endpoints therefore gives
\begin{equation}\label{eq:quantization_extreme}
    b_1(\lambda) = \frac{1}{2} + a_t + a_0 + n \,,
\end{equation}
where $n = \ell - |m_0| \ge 0$ represents the number of angular nodes. On the physical ordinary-point locus, $b_1=b_2=b$ is fixed by this condition and $u$, hence $\lambda$, is evaluated directly from \eqref{eq:reduced_matone}. The prepotential is truncated at this fixed momentum.

If $N$ denotes total
degree in the original quiver couplings $(t,L)$, only $\kappa^j=t^jL^j$ with
$2j\le N$ survives on the ordinary-point locus. Thus adjacent orders $2j$
and $2j+1$ coincide. Table~\ref{tab:extreme_cmetric} gives orders $j=1,2,3,4$ in $\kappa$ (equivalently $N=2,4,6,8$) and independent angular Galerkin benchmarks, using the same fixed-momentum prescription as Table~\ref{tab:generic_cmetric}.

\begin{table}[htbp]
\centering\footnotesize
\setlength{\tabcolsep}{3pt}\renewcommand{\arraystretch}{1.18}
\resizebox{\textwidth}{!}{\begin{tabular}{ccccccc}\toprule
$(m_0,\ell)$ & $j=1$ & $j=2$ & $j=3$ & $j=4$ & Numerical & $|\delta\lambda_{j=4}|$\\
\midrule
\multicolumn{7}{c}{$\alpha M=0.05$} \\
\midrule
$(1,1)$ & 2.6529724409 & 2.6529785952 & 2.6529785952 & 2.6529785952 & 2.6529785952 & $\lesssim 10^{-12}$ \\
$(1,2)$ & 6.8640044068 & 6.8640088883 & 6.8640088883 & 6.8640088883 & 6.8640088883 & $\lesssim 10^{-12}$ \\
$(2,2)$ & 7.3983770421 & 7.3984032952 & 7.3984032956 & 7.3984032956 & 7.3984032956 & $\lesssim 10^{-12}$ \\
$(2,3)$ & 13.8254024854 & 13.8254233649 & 13.8254233648 & 13.8254233648 & 13.8254233648 & $\lesssim 10^{-12}$ \\
\midrule
\multicolumn{7}{c}{$\alpha M=0.10$} \\
\midrule
$(1,1)$ & 3.0135164835 & 3.0136369465 & 3.0136369576 & 3.0136369575 & 3.0136369575 & $\lesssim 10^{-12}$ \\
$(1,2)$ & 7.4619968245 & 7.4620878058 & 7.4620877986 & 7.4620877986 & 7.4620877986 & $\lesssim 10^{-12}$ \\
$(2,2)$ & 8.5959363958 & 8.5964435139 & 8.5964436252 & 8.5964436242 & 8.5964436242 & $2.34\times10^{-12}$ \\
$(2,3)$ & 15.5121954724 & 15.5126158189 & 15.5126157794 & 15.5126157791 & 15.5126157791 & $2.28\times10^{-12}$ \\
\midrule
\multicolumn{7}{c}{$\alpha M=0.30$} \\
\midrule
$(1,1)$ & 4.8780168776 & 4.8974823640 & 4.8977264573 & 4.8977190656 & 4.8977190757 & $1.02\times10^{-8}$ \\
$(1,2)$ & 10.6729976151 & 10.6934119888 & 10.6933304195 & 10.6933318872 & 10.6933318609 & $2.63\times10^{-8}$ \\
$(2,2)$ & 14.6676029963 & 14.7392302828 & 14.7408785737 & 14.7408352687 & 14.7408343360 & $9.33\times10^{-7}$ \\
$(2,3)$ & 24.3565320248 & 24.4378034048 & 24.4379307210 & 24.4379175656 & 24.4379183497 & $7.84\times10^{-7}$ \\
\bottomrule\end{tabular}}
\caption{Extreme angular spectra ($Q=M$). The last column gives
the absolute error at $j=4$; differences below $10^{-12}$ are
reported conservatively at the numerical comparison floor.}
\label{tab:extreme_cmetric}
\end{table}

Note that when $m_0 = 0$, the equation and its corresponding parameter dictionary are significantly simplified. In this case we recover the known exact spectrum \cite{Kofron:2015separability}. Appendix~\ref{app:l0_proof} gives the quiver derivation and an independent Legendre reduction:
\begin{equation}
    \lambda_{exact} = \ell(\ell+1)(1-\alpha^2 M^2) + \frac{1}{3}(1-\alpha^2 M^2) \,.
\end{equation}

\subsection{Continuation towards Extremality and Larger Acceleration}

At fixed $a=\alpha M<1$, $P(x)$ varies smoothly and remains positive on the angular interval as $Q/M\to1$, while the two singularities outside this interval coalesce. We follow the same angular mode through this limit using the regular and confluent descriptions. Table~\ref{tab:near_extreme_cmetric} follows the mode $(m_0,\ell)=(2,2)$ at $a=\alpha M=0.30$ as $Q/M$ approaches one. The rows with $Q<M$ use the reduced Heun expansion, while the last row is computed independently from the confluent prepotential. The three columns use single-node degrees $6,10,14$, denoted by $k$ for the regular sum and $j$ for the confluent sum. The angular benchmarks approach the extreme value $14.7408343360$ continuously. At $Q/M=0.90$, increasing the degree from $6$ to $14$ reduces the absolute error from $8.57\times10^{-2}$ to $1.18\times10^{-4}$; the higher-charge rows reach substantially smaller errors at the same degree.

\begin{table}[htbp]
\centering\footnotesize
\setlength{\tabcolsep}{4pt}\renewcommand{\arraystretch}{1.18}
\resizebox{\textwidth}{!}{\begin{tabular}{cccccc}\toprule
$Q/M$ & Order $6$ & Order $10$ & Order $14$ & Numerical & $|\delta\lambda_{14}|$\\
\midrule
$0.9000$ & 14.4233060473 & 14.5057933564 & 14.5088497117 & 14.5089679002 & $1.18\times10^{-4}$ \\
$0.9900$ & 14.7163282198 & 14.7164738865 & 14.7164740098 & 14.7164740099 & $1.04\times10^{-10}$ \\
$0.9990$ & 14.7383862520 & 14.7383864903 & 14.7383864903 & 14.7383864903 & $\lesssim10^{-12}$ \\
$0.9999$ & 14.7405894317 & 14.7405894332 & 14.7405894332 & 14.7405894332 & $\lesssim10^{-12}$ \\
$1$ & 14.7408343389 & 14.7408343360 & 14.7408343360 & 14.7408343360 & $\lesssim10^{-12}$ \\
\bottomrule\end{tabular}}
\caption{Continuation towards $Q=M$ for $\alpha M=0.30$ and $(m_0,\ell)=(2,2)$. The instanton approximations are evaluated at the indicated degrees, with $k$ used for $Q<M$ and $j$ for $Q=M$. The last column compares degree fourteen with the independent angular spectrum.}
\label{tab:near_extreme_cmetric}
\end{table}

In the extreme frame, $t=-2a/(1-a)$ and $L=-|m_0|t$, so the reduced instanton coordinate is
\begin{equation}
 \kappa=tL=-\frac{4|m_0|a^2}{(1-a)^2}.
\end{equation}
The bound $|t|<1$ applies to the original sewing coordinate and corresponds to $a<1/3$. After the ordinary point is removed, the series depends on $\kappa$ alone. Table~\ref{tab:extended_extreme_cmetric} evaluates this reduced series at $a=0.40$, where $t=-4/3$, for four nonaxisymmetric modes. The errors decrease across the displayed degrees; for $(m_0,\ell)=(1,1)$, they fall from $9.97\times10^{-6}$ at $j=4$ to $3.08\times10^{-12}$ at $j=10$.

\begin{table}[htbp]
\centering\footnotesize
\setlength{\tabcolsep}{3pt}\renewcommand{\arraystretch}{1.18}
\resizebox{\textwidth}{!}{\begin{tabular}{ccccccc}\toprule
$(m_0,\ell)$ & $j=4$ & $j=6$ & $j=8$ & $j=10$ & Numerical & $|\delta\lambda_{10}|$\\
\midrule
$(1,1)$ & 6.1084961847 & 6.1084861868 & 6.1084862118 & 6.1084862119 & 6.1084862119 & $3.08\times10^{-12}$ \\
$(1,2)$ & 12.8901606433 & 12.8901683399 & 12.8901683081 & 12.8901683079 & 12.8901683079 & $3.43\times10^{-12}$ \\
$(2,2)$ & 18.6023348803 & 18.6022323151 & 18.6022332817 & 18.6022332691 & 18.6022332693 & $1.64\times10^{-10}$ \\
$(2,3)$ & 30.2574960006 & 30.2575267453 & 30.2575257975 & 30.2575258115 & 30.2575258113 & $1.79\times10^{-10}$ \\
\bottomrule\end{tabular}}
\caption{Extreme angular spectra at $\alpha M=0.40$. The column labels give the instanton degree $j$ in $\kappa$.}
\label{tab:extended_extreme_cmetric}
\end{table}

Appendix~\ref{app:numerics} specifies the regulator, precision and angular refinement checks for these comparisons.

\section{Conclusion and Outlook}
\label{sec:conclusion}

In this paper, we develop the angular quiver description of the charged C-metric and evaluate its extreme spectrum. The regular five-point system and its confluent limit are connected by a hypermultiplet decoupling prescription, with accessory parameters that remain finite after the heavy-mass terms are removed. An explicit coordinate transformation relates this construction to the physical limit $Q\to M$. Regularity at the two angular poles gives an algebraic quantization condition through the connection formula. Comparisons with independent angular spectra test the resulting instanton approximations, including the approach to extremality and the reduced expansion at $\alpha M=0.40$. The axisymmetric sector recovers the exact spectrum of \cite{Kofron:2015separability}.

Building on these results, several directions remain for future work. First, this confluent gauge theory framework can be extended to analyze wave equations and boundary value problems (such as quasinormal modes or superradiance) in black holes with more complex geometric and horizon structures. Second, one can investigate more severe singularity collision limits, such as doubly confluent or multiply confluent setups. These correspond to higher-rank irregular states in the conformal field theory \cite{Bonelli:2022ten}. Finally, the decoupling construction can be studied for quiver topologies beyond linear chains, including affine or D/E-type star-shaped quivers \cite{Gaiotto:2009we, Bao:2011rc}.
\vspace{8pt}

$\\$
\noindent {\it Acknowledgements.} 
The authors thank  Xian-Hui Ge, Masataka Matsumoto, Yutaka Matsuo, Jean-Emile Bourgine, Futoshi Yagi, Yang Lei, Hongfei Shu, Rui-Dong Zhu and Yi-Rong Wang
for helpful discussions.  K.Z. (Hong Zhang) is supported by a classified fund of Shanghai city.

\appendix

\section{Exact Prepotential and its Confluent Limit}
\label{app:combinatorics}

In this appendix, we provide the complete combinatorial definition of the Nekrasov-Shatashvili (NS) prepotential. We unify the expressions for the generic $\mathrm{SU(2)} \times \mathrm{SU(2)}$ quiver used in Section \ref{sec:generic_c_metric} and formally generalize the exact decoupling framework to arbitrary $\mathrm{SU(2)}^{n-3}$ linear quiver gauge theories. This generalized framework gives canonical equations with $n-2$ regular singular points and one rank-1 irregular singularity; the physical C-metric occupies a special removable-puncture locus.

For computational convenience, the full Nekrasov-Shatashvili prepotential is decomposed as
\begin{equation}
    \mathcal{F}^{\text{full}} = F_{\text{tree}}^{\text{Coulomb}} + F^{\text{NS}} \,,
\end{equation}
where $F^{\text{NS}}$ denotes the tree-subtracted prepotential adopted in the main text,
\begin{equation}\label{eq:F_NS_decomp}
    F^{\text{NS}} = F_{\text{1-loop}} + F_{\text{q}} \,, \quad \text{with} \quad F_{\text{q}} \equiv F_{U(1)} + F_{\text{inst}} \,.
\end{equation}

\subsection{The Generic Quiver Prepotential}

The instanton partition function is formulated as a sum over arrays of Young diagrams \cite{Nekrasov:2002qd,Nekrasov:2003rj}. For pure $\mathrm{SU(2)}$ theory, instanton recursion relations also provide a geometric construction of the multi-instanton measure \cite{Matone:1995rx, Matone:2001measure}. Let $Y$ be a Young diagram with row lengths $\lambda_i$ and column heights $\lambda'_j$. For any box $s = (i,j) \in \mathbb{Z}_{\ge 1}^2$, the arm length $A_Y(s)$ and the leg length $L_Y(s)$ are defined as $A_Y(i, j) = \lambda_i - j$ and $L_Y(i, j) = \lambda'_j - i$.

To define the partition function in the NS limit ($\epsilon_1 = 1, \epsilon_2 \to 0$), we introduce the fundamental algebraic building blocks. For a pair of Young diagrams $\vec{Y} = (Y_1, Y_2)$ and a doublet of Coulomb branch parameters $\vec{b} = (b, -b)$, we define the auxiliary function:
\begin{equation}
    E(x, Y_k, Y_l, s) = x - L_{Y_l}(s) + \epsilon_2 \left( A_{Y_k}(s) + 1 \right) \,.
\end{equation}
The contributions of the hypermultiplet, vector multiplet, and bifundamental multiplet are defined respectively as:
\begin{align}
    z_{\text{hyp}} \left( \vec{b}, \vec{Y}, m \right) &= \prod_{k=1,2} \prod_{(i,j) \in Y_k} \left( b_k + m + i - \frac{1}{2} + \epsilon_2 \left( j - \frac{1}{2} \right) \right) \,, \label{eq:app_zhyp} \\
    z_{\text{vec}} \left( \vec{b}, \vec{Y} \right) &= \prod_{k,l=1,2} \prod_{(i,j) \in Y_k} E^{-1} \left( b_k - b_l, Y_k, Y_l, (i, j) \right) \nonumber \\
    &\quad \times \prod_{(i',j') \in Y_l} \left( 1 + \epsilon_2 - E \left( b_l - b_k, Y_l, Y_k, (i', j') \right) \right)^{-1} \,, \label{eq:app_zvec} \\
    z_{\text{bif}} \left( \vec{b}_1, \vec{Y}, \vec{b}_2, \vec{W}; m \right) &= \prod_{k,l=1,2} \prod_{(i,j) \in Y_k} \left( E \left( (b_1)_k - (b_2)_l, Y_k, W_l, (i, j) \right) - \left( \frac{1+\epsilon_2}{2} + m \right) \right) \nonumber \\
    &\quad \times \prod_{(i',j') \in W_l} \left( \left( \frac{1+\epsilon_2}{2} - m \right) - E \left( (b_2)_l - (b_1)_k, W_l, Y_k, (i', j') \right) \right) \,. \label{eq:app_zbif}
\end{align}

For the regular 5-point setup applied in Section \ref{sec:generic_c_metric}, the generic singular points are strictly located at $\{0, t, 1, q, \infty\}$. The full NS prepotential evaluated on this geometry is decomposed as:
\begin{equation}\label{eq:fullns_generic}
    \mathcal{F}^{\text{full}}\left(1/q, t\right) = F_{\text{tree}}^{\text{Coulomb}} + F_{U(1)} + F_{\text{inst}}\left(1/q, t\right) + F_{\text{1-loop}} \,.
\end{equation}

The instanton partition function is assembled directly from the combinatorial blocks:
\begin{equation}
\begin{aligned}
    Z_{\text{inst}} &= \sum_{\vec{Y}, \vec{W}} t^{|\vec{Y}|} \left(\frac{1}{q}\right)^{|\vec{W}|} z_{\text{vec}} \left( \vec{b}_1, \vec{Y} \right) z_{\text{vec}} \left( \vec{b}_2, \vec{W} \right) z_{\text{bif}} \left( \vec{b}_1, \vec{Y}, \vec{b}_2, \vec{W}; -a_1 \right) \\
    &\quad \times \left( \prod_{\sigma = \pm 1} z_{\text{hyp}} \left( \vec{b}_1, \vec{Y}, a_t + \sigma a_0 \right) \right) \left( \prod_{\sigma = \pm 1} z_{\text{hyp}} \left( \vec{b}_2, \vec{W}, a_q + \sigma a_\infty \right) \right) \,.
\end{aligned}
\end{equation}
The instanton prepotential $F_{\text{inst}}$ is then extracted via the standard $\Omega$-deformation limit $F_{\text{inst}}(1/q, t) = \lim_{\epsilon_2 \to 0} \epsilon_2 \ln Z_{\text{inst}}(1/q, t)$.
The $U(1)$ normalization term $F_{U(1)}$ comes from the pair factors of the regular conformal block and is a sum of logarithms:
\begin{equation}
\begin{aligned}
F_{U(1)}
&= -2\left(a_1+\frac{1}{2}\right)\left(a_t+\frac{1}{2}\right)\ln(1-t) \\
&\quad - 2\left(\frac{1}{2}-a_1\right)\left(a_q+\frac{1}{2}\right)\ln\left(1-\frac{1}{q}\right) \\
&\quad - 2\left(a_t+\frac{1}{2}\right)\left(a_q+\frac{1}{2}\right)\ln\left(1-\frac{t}{q}\right) \,.
\end{aligned}
\end{equation}
The one-loop prepotential is expressed through a primitive $\psi^{(-2)}(x)$ defined by $\partial_x\psi^{(-2)}(x)=\log\Gamma(x)$ and $\psi^{(-2)}(1)=0$, with analytic continuation on the chosen branch:
\begin{equation}
\begin{aligned}
    F_{\text{1-loop}} &= \sum_{\theta=\pm 1} \left( \psi^{(-2)}(1+2\theta b_1) + \psi^{(-2)}(1+2\theta b_2) \right) \\ &\quad- \sum_{\theta_1,\theta_2=\pm 1} \psi^{(-2)}\left(\frac{1}{2} + a_1 + \theta_1 b_1 + \theta_2 b_2\right) \\
    &\quad - \sum_{\theta,\sigma=\pm 1} \psi^{(-2)}\left(\frac{1}{2} + \theta b_1 - a_t + \sigma a_0\right) - \sum_{\theta,\sigma=\pm 1} \psi^{(-2)}\left(\frac{1}{2} + \theta b_2 - a_q + \sigma a_\infty\right) \,.
\end{aligned}
\end{equation}
Finally, the purely classical Coulomb branch term reads:
\begin{equation}
    F_{\text{tree}}^{\text{Coulomb}} = -b_1^2\,\ln(t) - b_2^2\,\ln(1/q) \,.
\end{equation}
Crucially, unlike $F_{U(1)}$, $F_{\text{inst}}$, or $F_{\text{1-loop}}$, this term does not expand into a power series of the geometrical cross-ratios. Instead, it serves as the bare vacuum background carrying the internal momenta $b_i$. 
In \eqref{eq:connection_formula} the one-loop contribution has already been converted to its explicit Gamma factors. The remaining derivatives act on $F_{\rm q}$ in the fixed mass convention of the instanton sum, at fixed internal Coulomb parameters; no implicit mass substitution is made.

\subsection{The Generalized Confluent Prepotential}

Consider the collision limit of an $n$-point regular BPZ equation, producing a rank-1 irregular singularity at $z=\infty$. To unify the notation for the remaining $n-2$ regular punctures, we denote their positions as $\{0, z_1, z_2, \dots, z_{n-4}, 1\}$, and assign $z_{n-3} \equiv 1$ with the corresponding mass parameter $a_{z_{n-3}} \equiv a_1$. 

Put $r=n-3$ and choose the nested channels surrounding
$\{0,z_1,\ldots,z_k\}$, with
$0<|z_1|<\cdots<|z_{r-1}|<z_r=1$.
The corresponding sewing parameters are
\begin{equation}\label{eq:sewing}
 \mathsf p_k=\frac{z_k}{z_{k+1}}\quad(k=1,\ldots,r-1),
 \qquad \mathsf p_r=L.
\end{equation}
Before confluence the last coupling is $1/q$; its heavy-mass limit produces
the weight $L^{|\vec Y_r|}$. The geometric ordering ensures
$|\mathsf p_k|<1$ for $k<r$. The instanton sum is expanded near $L=0$,
with $L$ measured in $\epsilon_1=1$ units.

The decoupling limit of the two outermost punctures, initially at $q$ and $\infty$, is defined by
\begin{equation}
    a_q = \frac{\eta - m_{SW}}{2} \,, \quad a_\infty = \frac{-\eta - m_{SW}}{2} \,, \quad q = \frac{\eta}{L} \,, \quad \text{with} \quad \eta \to \infty \,.
\end{equation}
At fixed Young diagrams, the heavy hypermultiplet factor scales as
$z_{\text{hyp}}(\vec b_r,\vec Y_r,\eta)\sim\eta^{|\vec Y_r|}$, so
\begin{equation}
 \left(\frac{L}{\eta}\right)^{|\vec Y_r|}
 z_{\text{hyp}}(\vec b_r,\vec Y_r,\eta)\longrightarrow L^{|\vec Y_r|}.
\end{equation}
The other right-node hypermultiplet factor retains mass $-m_{SW}$.
The instanton partition function for the generalized confluent
$\mathrm{SU(2)}^{n-3}$ quiver theory reads:
\begin{equation}
\begin{aligned}
    Z_{\text{inst}}^{\text{conf}} &= \sum_{\vec{Y}_1, \dots, \vec{Y}_{n-3}} \left( \prod_{k=1}^{n-4} (\mathsf{p}_k)^{|\vec{Y}_k|} \right) L^{|\vec{Y}_{n-3}|} \left( \prod_{j=1}^{n-3} z_{\text{vec}}(\vec{b}_j, \vec{Y}_j) \right) \\
    &\quad \times \left( \prod_{j=1}^{n-4} z_{\text{bif}}(\vec{b}_j, \vec{Y}_j, \vec{b}_{j+1}, \vec{Y}_{j+1}; -a_{z_{j+1}}) \right) \\
    &\quad \times \left( \prod_{\sigma=\pm} z_{\text{hyp}}(\vec{b}_1, \vec{Y}_1, a_{z_1} + \sigma a_0) \right) \times z_{\text{hyp}}(\vec{b}_{n-3}, \vec{Y}_{n-3}, -m_{SW}) \,.
\end{aligned}
\end{equation}
The instanton prepotential is extracted via $F_{\text{inst}}^{\text{conf}} = \lim_{\epsilon_2 \to 0} \epsilon_2 \ln Z_{\text{inst}}^{\text{conf}}$. Note that setting $n=5$ with $z_1 = t$ and $z_2 = 1$ reproduces the two-node confluent instanton sum used in Section \ref{sec:confluent_quiver}.

The $U(1)$ normalization is obtained from the oriented pair factors of the
regular block \cite{Alday:2009aq,Arnaudo:2025kof}, rather than from the classical
Coulomb term. Define
\begin{equation}
 c_1=\frac12+a_{z_1},\qquad
 c_i=\frac12-a_{z_i}\quad(2\le i\le r).
\end{equation}
Then its confluent limit in the bifundamental convention above is
\begin{equation}\label{eq:general_u1}
 F_{U(1)}^{\rm conf}
 =-2\sum_{1\le i<j\le r}c_i\left(a_{z_j}+\frac12\right)
       \log\left(1-\frac{z_i}{z_j}\right)
   +L\sum_{i=1}^r c_i z_i.
\end{equation}
For $r=2$, this reproduces the two-node $U(1)$ term in the main text.
The signs of the interior coefficients depend on the orientation of the
bifundamental building blocks and must be changed together with that convention.

In the one-loop prepotential, the right-node hypermultiplet terms have
arguments $1/2+\theta b_r-\eta$ and $1/2+\theta b_r+m_{SW}$.
We absorb the heavy hypermultiplet contribution into the finite part,
keeping the remaining parameters fixed and using the same analytic branch:
\begin{equation}\label{eq:one_loop_finite_part}
 F_{\text{1-loop}}^{\text{conf}}
 =\lim_{\eta\to\infty}\left[
 F_{\text{1-loop}}^{\text{reg}}
 +\sum_{\theta=\pm}\psi^{(-2)}
       \left(\frac12+\theta b_r-\eta\right)\right].
\end{equation}
The resulting one-loop prepotential is
\begin{equation}
\begin{aligned}
    F_{\text{1-loop}}^{\text{conf}} &= \sum_{j=1}^{n-3} \sum_{\theta=\pm} \psi^{(-2)}(1 + 2\theta b_j) - \sum_{j=1}^{n-4} \sum_{\theta_1, \theta_2 = \pm} \psi^{(-2)}\left(\frac{1}{2} + a_{z_{j+1}} + \theta_1 b_j + \theta_2 b_{j+1}\right) \\
    &\quad - \sum_{\theta, \sigma = \pm} \psi^{(-2)}\left(\frac{1}{2} - a_{z_1} + \theta b_1 + \sigma a_0\right) - \sum_{\theta = \pm} \psi^{(-2)}\left(\frac{1}{2} + \theta b_{n-3} + m_{SW}\right) \,.
\end{aligned}
\end{equation}

The confluent BPZ equation is
\begin{equation}\label{eq:gen_confluent_bpz}
\begin{aligned}
    \Bigg[ \frac{\mathrm{d}^2}{\mathrm{d}z^2} &+ \frac{\frac{1}{4}-a_0^2}{z^2} + \sum_{k=1}^{n-4}\frac{\frac{1}{4}-a_{z_k}^2}{(z-z_k)^2} + \frac{\frac{1}{4}-a_1^2}{(z-1)^2} \\
    &+ \frac{u_0}{z(z-1)} + \sum_{k=1}^{n-4}\frac{(z_k-1)u_{z_k}}{z(z-1)(z-z_k)} - \frac{L^2}{4} + \frac{m_{SW} L}{z} \Bigg] \psi(z) = 0 \,.
\end{aligned}
\end{equation}
The internal Coulomb momenta are related to the accessory parameters by the renormalized Matone relations. For $k=1,\ldots,n-4$, they read
\begin{equation}
    b_k^2 = - \frac{k}{4} + a_0^2 + \sum_{j=1}^k a_{z_j}^2 - \sum_{j=1}^k u_{z_j} + \sum_{j=1}^k z_j \partial_{z_j} F_{\rm q}^{\text{conf}} \,,
\end{equation}
For the outermost Coulomb momentum $b_{n-3}$, the corresponding relation is
\begin{equation}
    b_{n-3}^2 = - \frac{n-3}{4} + a_0^2 + \sum_{j=1}^{n-3} a_{z_j}^2 - u_0 + L \partial_L F_{\rm q}^{\text{conf}} \,,
\end{equation}
where $F_{\rm q}^{\text{conf}}=F_{\text{inst}}^{\text{conf}}+F_{U(1)}^{\text{conf}}$ is the normalized confluent instanton prepotential, $a_{z_{n-3}} \equiv a_1$, and the regulated accessory parameter $u_0$ is intrinsically defined by isolating the diverging bare scaling dimensions $\Delta_i \equiv 1/4 - a_i^2$:
\begin{equation}
    u_0 \equiv \lim_{\eta \to \infty} \left[ a_q^2 - a_\infty^2 - \left(\Delta_0 + \sum_{j=1}^{n-3} \Delta_{z_j}\right) - u_q \right].
\end{equation}
For $n=5$, the internal and outermost relations reproduce Eq.\eqref{eq:renormalized_matone}.

The differential operators in the internal Matone relations satisfy
\begin{equation}
 \sum_{j=1}^k z_j\partial_{z_j}=\mathsf p_k\partial_{\mathsf p_k},
 \qquad k<r,
\end{equation}
at fixed remaining sewing parameters. The one-loop term is independent of these
parameters, so only $F_{\rm q}=F_{\rm inst}+F_{U(1)}$ contributes to these derivatives.

\paragraph{Six-point consistency check.}
For $n=6$ there are three nodes, with
$\mathsf p_1=z_1/z_2$, $\mathsf p_2=z_2$, and $\mathsf p_3=L$.
The bifundamental masses are $-a_{z_2}$ and $-a_1$, and
\begin{equation}
\begin{aligned}
 F_{U(1)}^{\rm conf}
 &=-2(a_{z_1}+\tfrac12)(a_{z_2}+\tfrac12)\log(1-\mathsf p_1)\\
 &\quad-2(a_{z_1}+\tfrac12)(a_1+\tfrac12)\log(1-\mathsf p_1\mathsf p_2)\\
 &\quad-2(\tfrac12-a_{z_2})(a_1+\tfrac12)\log(1-\mathsf p_2)\\
 &\quad+L\left[(a_{z_1}+\tfrac12)\mathsf p_1\mathsf p_2
             +(\tfrac12-a_{z_2})\mathsf p_2+\tfrac12-a_1\right].
\end{aligned}
\end{equation}
Writing $\Delta_x=1/4-x^2$, the one-box sums give
\begin{equation}
\begin{aligned}
 F_{\rm q}^{\rm conf}&=C_1\mathsf p_1+C_2\mathsf p_2+C_3L+O(\mathsf p^2),\\
 C_1&=\frac{(\Delta_{b_1}+\Delta_{a_{z_1}}-\Delta_{a_0})
                 (\Delta_{b_1}+\Delta_{a_{z_2}}-\Delta_{b_2})}
                 {2\Delta_{b_1}},\\
 C_2&=\frac{(\Delta_{b_2}+\Delta_{a_{z_2}}-\Delta_{b_1})
                 (\Delta_{b_2}+\Delta_{a_1}-\Delta_{b_3})}
                 {2\Delta_{b_2}},\\
 C_3&=\frac{m_{\rm SW}(4a_1^2+4b_3^2-4b_2^2-1)}{2(4b_3^2-1)}.
\end{aligned}
\end{equation}
Here $O(\mathsf p^2)$ denotes total degree at least two in
$(\mathsf p_1,\mathsf p_2,L)$. The first two coefficients equal the level-one
Virasoro sewing coefficients, while the third is the confluent limit of the
rightmost regular-node coefficient. This checks the orientation and sewing
conventions beyond the two-node specialization.

\subsection{Prepotentials with Vacuum Insertions}
\label{app:physical_reductions}

\paragraph{Regular angular equation.}
The four-point coordinate used to evaluate the spectrum in
Section~\ref{sec:generic_c_metric} is
\begin{equation}\label{eq:heun_reduction}
 w=\frac{(q-1)z}{q-z},\qquad
 T=\frac{t(1-p)}{1-pt},\qquad p=\frac1q,
\end{equation}
which sends $0,t,1,q$ to $0,T,1,\infty$. With the normal-form wavefunction
transformation, its endpoint masses are $a_0,a_t,a_1,a_q$, and its accessory
parameter is given by \eqref{eq:heun_accessory}.
Write $\Delta_b=1/4-b^2$ and denote the regular two-node prepotential by $F_5(t,p;b_1,b_2)=F_{\rm q}(p,t)$. On the vacuum locus $a_\infty=1/2$, $b_1=b$ and $b_2=a_q$, its normalization is related to \eqref{eq:reduced_heun_F} by
\begin{equation}\label{eq:regular_prepotential_reduction}
\begin{aligned}
 F_5(t,p;b,a_q)&=\mathcal F_4(T;b)
 -(\Delta_q+\Delta_1-\Delta_b)\log(1-p)\\
 &\quad-(\Delta_b-\Delta_0+\Delta_t)\log(1-pt).
\end{aligned}
\end{equation}
The logarithms are the Jacobian factors of \eqref{eq:heun_reduction}. To see their normalization explicitly, set $B=\Delta_b-\Delta_0+\Delta_t$. The transformed Matone relation gives
\begin{equation}
 t\partial_tF_5
 =\frac{T\partial_T\mathcal F_4+ptB}{1-pt}.
\end{equation}
Integrating at fixed $p$ fixes the second logarithm in \eqref{eq:regular_prepotential_reduction}; the three-point block remaining at $t=0$ has normalized value
$F_5(0,p)=-(\Delta_q+\Delta_1-\Delta_b)\log(1-p)$, fixing the first. The same relation satisfies the second Matone equation and \eqref{eq:ordinary_constraint}. Thus the two-node and reduced descriptions give the same accessory parameter.

\paragraph{Extreme angular equation.}
The physical values $a_1^2=1/4$ and $u_t=u_0=u$ remove the marked point
$z=1$ identically:
\begin{equation}
 \frac{u}{z(z-1)}+\frac{(t-1)u}{z(z-1)(z-t)}=\frac{u}{z(z-t)}.
\end{equation}
With $w=z/t$ and $\kappa=tL$, the resulting standard confluent Heun equation is
\begin{equation}\label{eq:reduced_che}
 \left[\partial_w^2+\frac{1/4-a_0^2}{w^2}
 +\frac{1/4-a_t^2}{(w-1)^2}+\frac{u}{w(w-1)}
 -\frac{\kappa^2}{4}+\frac{m_{\rm SW}\kappa}{w}\right]\psi=0.
\end{equation}
The physical interval is $0\le w\le1$. Before taking the NS limit, the
removed point is represented by the zero-weight insertion with
$a_1=(1+\varepsilon)/2$. Removing this vacuum insertion identifies the adjacent
internal modules; we choose $b_1=b_2=b$ on the weak-coupling branch.
The $O(\varepsilon)$ difference between this mass and $a_1=1/2$ does not change
the coefficientwise NS prepotential at generic Coulomb parameters.
This gives \eqref{eq:reduced_prepotential} as a formal weak-coupling
series, with resonant values reached by continuation.

The single-node instanton partition function is
\begin{equation}\label{eq:reduced_partition}
 Z_3(\kappa)=\sum_{\vec Y}\kappa^{|\vec Y|}z_{\rm vec}(\vec b,\vec Y)
 \prod_{\sigma=\pm}z_{\rm hyp}(\vec b,\vec Y,a_t+\sigma a_0)
 z_{\rm hyp}(\vec b,\vec Y,-m_{\rm SW}).
\end{equation}
The same reduction can be followed directly in the regular collision frame
\eqref{eq:physical_collision_frame}. Removing its ordinary point and scaling
$\xi=\widehat z/\widehat q$ gives a four-point equation with parameter
$\widehat T=\widehat t/\widehat q$. If $w_R$ denotes the four-point coordinate
$w$ in \eqref{eq:heun_reduction}, the two regular frames are related by
\begin{equation}\label{eq:reduced_coordinate_bridge}
 \xi=\frac{w_R}{w_R-1},\qquad \widehat T=\frac{T}{T-1}.
\end{equation}
The normalized prepotentials transform accordingly.\footnote{In the
normalization where both prepotentials vanish at zero coupling,
$\widehat{\mathcal F}_4(T/(T-1);b)
=\mathcal F_4(T;b)+(\Delta_b-\Delta_0+\Delta_t)\log(1-T)$.
The hat also changes the right-hand mass representatives from
$(a_1^{(R)},a_q^{(R)})$ to $(a_q^{(R)},-a_1^{(R)})$. The corresponding pairs
of hypermultiplet masses are $(\eta,\widehat m_{\rm SW})$ and
$(\eta,-\widehat m_{\rm SW})$, respectively.}
The right-node hypermultiplet masses of the reduced single-node sum are
$-\widehat m_{\rm SW}$ and $\eta$. At each instanton degree $j$, the factor $\widehat T^j z_{\rm hyp}(\vec b,\vec Y,\eta)$ tends to $\kappa^j$, since $\eta\widehat T=\widehat t\widehat L\to\kappa$. Its normalization satisfies
\begin{equation}
 -2(\widehat a_q+\tfrac12)(a_t+\tfrac12)\log(1-\widehat T)
 \longrightarrow(a_t+\tfrac12)\kappa.
\end{equation}
Thus the reduced $N_f=4$ sum decouples to \eqref{eq:reduced_partition}, with the normalization in \eqref{eq:reduced_prepotential}. The physical reduction and the quiver collision give the same single-node prepotential coefficient by coefficient.

As an explicit check,
\begin{equation}
 [\kappa]\mathcal F_3=
 \frac{2m_{\rm SW}(b^2+a_t^2-a_0^2-1/4)}{4b^2-1}.
\end{equation}
The derivatives $t\partial_t$ and $L\partial_L$ both act as
$\kappa\partial_\kappa$ on this prepotential, yielding
\eqref{eq:reduced_matone}.

\section{Analytical Eigenvalues for
\texorpdfstring{$m_0=0$}{m0=0} in the Extreme Limit}
\label{app:l0_proof}

The spectrum obtained below is the scalar specialization of the result in
\cite{Kofron:2015separability}. We first recover it in the quiver convention and
then give an independent differential-equation derivation.

In the axisymmetric sector, the extreme C-metric dictionary
\eqref{eq:ext_masses}--\eqref{eq:ext_u0_ut} reduces to
\begin{equation}
    a_0=a_t=0,\qquad a_1^2=\frac14,\qquad
    L=m_{\rm SW}=0,\qquad
    u_0=u_t=u\equiv
    -\frac16-\frac{\lambda}{1-\alpha^2M^2}.
\end{equation}
The numerical convention is $a_1=+1/2$.
For the combinatorial proof, we first use the reflected
representative $a_1=-1/2$ and then translate the result back
to this convention.
In the normalized conformal-block convention,
$F_{\rm q}=F_{\rm inst}+F_{U(1)}$ is invariant under
$a_1\mapsto-a_1$, since the corresponding conformal weight
depends only on $a_1^2$.
Its two contributions need not be separately invariant.

\paragraph{Vanishing NS corrections.}
At $L=0$, the weak-coupling expansion gives
$L\partial_LF_{\rm q}=0$.
The two renormalized Matone relations therefore become
\begin{equation}\label{eq:app_matone_pair}
    b_2^2=-\frac14-u,\qquad
    b_1^2-b_2^2=t\partial_tF_{\rm q}(t,0;b_1,b_2),
\end{equation}
where the derivative is taken at fixed Coulomb and mass parameters.

To evaluate the right-hand side, first consider $b_1=b_2=b$
at generic $2b\notin\mathbb Z$, with $\epsilon_1=1$ and
$\epsilon_2=\varepsilon$.
Setting $L=0$ restricts the second-node diagrams to
$\vec W=\varnothing$, while the sum over $\vec Y$ remains.
For each diagram $Y$, the same-colour vector factor has one
$\varepsilon^{-1}$ pole per column, arising from boxes with
$L_Y(s)=0$.
On the other hand, the same-colour bifundamental factor is
\begin{equation}
    E(0,Y,\varnothing,s)
    -\left(\frac{1+\varepsilon}{2}-a_1\right)
    =
    i-1+\varepsilon\left(A_Y(s)+\frac12\right),
    \qquad s=(i,j).
\end{equation}
It supplies one factor of $\varepsilon$ for every box in the
first row. Since the first-row length equals the number of
columns, these zeros balance all same-colour vector poles.
The remaining factors are finite at generic $b$.

Consequently, writing
$Z_{\rm inst}=1+\sum_{N\geq1}Z_N(\varepsilon)t^N$,
each coefficient satisfies $Z_N(\varepsilon)=O(1)$, and hence
\begin{equation}
    [t^N]F_{\rm inst}
    =
    \lim_{\varepsilon\to0}
    \varepsilon[t^N]\log Z_{\rm inst}
    =0,\qquad N\geq1.
\end{equation}
Moreover, $F_{U(1)}=0$ for $a_1=-1/2$ and $L=0$.
Thus, to all orders in the weak-coupling expansion,
\begin{equation}\label{eq:app_Fq_zero}
    F_{\rm q}(t,0;b,b)=0.
\end{equation}
For the reflected branch $a_1=-1/2$, the preceding argument
gives $F_{\rm inst}=0$, while $F_{U(1)}=0$ directly.
For the numerical branch $a_1=+1/2$, reflection invariance
instead implies, on the same locus $L=0$ and $b_1=b_2$,
\begin{equation}
    F_{U(1)}=-(1+2a_t)\log(1-t),
    \qquad
    F_{\rm inst}=(1+2a_t)\log(1-t).
\end{equation}
Thus $F_{\rm q}=0$ in either representation.
In particular, at $a_t=0$ the two contributions in the positive
branch are $-\log(1-t)$ and $\log(1-t)$, respectively.
The subsequent Matone and regulated connection conditions
therefore give the same spectrum in both branches.

Equation \eqref{eq:app_Fq_zero} shows that $b_1(t)=b_2$
solves \eqref{eq:app_matone_pair}.
It is the unique perturbative branch satisfying
$b_1(0)=b_2$ at generic $b_2$, so
\begin{equation}\label{eq:app_b1_exact}
    b_1^2=b_2^2=-\frac14-u.
\end{equation}
The resulting identities are continued to resonant Coulomb
parameters only after this simplification.

\paragraph{Resonant connection condition.}
Let $\psi_{0,\pm}$ and $\psi_{t,+}$ denote the locally normalized
Frobenius solutions used in \eqref{eq:connection_formula}, and define
$C_\pm$ by
\begin{equation}
    \psi_{t,+}=C_+\psi_{0,+}+C_-\psi_{0,-}.
\end{equation}
Introduce $a_0=\delta$ and temporarily keep $a_t$ nonzero.
Using \eqref{eq:app_Fq_zero} and its mass derivatives,
the connection coefficients reduce to
\begin{equation}
    C_\sigma=
    e^{i\pi(\frac12+a_t)}t^{a_t-\sigma\delta}
    \frac{\Gamma(1+2a_t)\Gamma(-2\sigma\delta)}
    {\Gamma(\frac12+a_t-\sigma\delta+b_1)
     \Gamma(\frac12+a_t-\sigma\delta-b_1)},
    \qquad \sigma=\pm1.
\end{equation}

As $\delta\to0$, the two solutions coalesce.
Define the regular and logarithmic solutions by
\begin{equation}
    \psi_{0,\rm reg}
    =\lim_{\delta\to0}\frac{\psi_{0,+}+\psi_{0,-}}2,
    \qquad
    \psi_{0,\log}
    =\lim_{\delta\to0}
      \frac{\psi_{0,+}-\psi_{0,-}}{2\delta},
\end{equation}
where $\psi_{0,\log}\sim z^{1/2}\log z$.
The change of basis gives
\begin{equation}
    C_+\psi_{0,+}+C_-\psi_{0,-}
    =
    (C_++C_-)\frac{\psi_{0,+}+\psi_{0,-}}2
    +\delta(C_+-C_-)
     \frac{\psi_{0,+}-\psi_{0,-}}{2\delta}.
\end{equation}
Thus regularity at both physical poles requires the logarithmic
connection coefficient to vanish:
\begin{equation}\label{eq:app_resonant_condition}
    D_{\log}
    \equiv
    \lim_{a_t\to0}\lim_{\delta\to0}
    \delta(C_+-C_-)=0.
\end{equation}
Using $\delta\Gamma(\mp2\delta)\to\mp1/2$, we obtain
\begin{equation}
    D_{\log}
    =
    -\frac{i}
    {\Gamma(\frac12+b_1)\Gamma(\frac12-b_1)}
    =
    -\frac{i}{\pi}\cos(\pi b_1).
\end{equation}
An overall phase depends on the local branch conventions and
does not affect its zeros.
Choosing the positive Weyl branch yields
\begin{equation}
    b_1=\ell+\frac12,\qquad \ell\in\mathbb Z_{\geq0}.
\end{equation}
Combining this condition with \eqref{eq:app_b1_exact} and the
geometric dictionary gives
\begin{equation}
    \lambda_{\rm exact}
    =(1-\alpha^2M^2)
    \left[\ell(\ell+1)+\frac13\right].
\end{equation}

\paragraph{Independent Legendre reduction.}
Set $a=\alpha M$ and, for $m_0=0$, introduce
\begin{equation}
 s=\frac{x-a}{1-ax},\qquad \chi(x)=\frac{f(s)}{1-ax}.
\end{equation}
Substitution into \eqref{eq:normal_potential} with
$H=(1-x^2)(1-ax)^2$ gives
\begin{equation}
 (1-s^2)f''-2sf'+\left[\frac{\lambda}{1-a^2}-\frac13\right]f=0.
\end{equation}
For $0\le a<1$, this transformation maps $[-1,1]$ monotonically onto itself,
with nonvanishing prefactor. Regularity at both endpoints therefore gives
\begin{equation}
 \chi_\ell(x)\propto\frac{P_\ell((x-a)/(1-ax))}{1-ax},\qquad
 \lambda_\ell=(1-a^2)\left[\ell(\ell+1)+\frac13\right].
\end{equation}
The result holds throughout $0\le a<1$, independently of the convergence
domain of a chosen instanton coordinate. This also exhibits explicitly the
hypergeometric degeneration of the axisymmetric equation.

\section{Numerical Methods and Reproducibility}
\label{app:numerics}
\paragraph{Angular eigenvalues.}
The benchmark computation uses the self-adjoint form
\begin{equation}
 -(H\chi')'+\left(\frac{m^2}{H}-\frac{H''}{6}\right)\chi=\lambda\chi,
 \qquad H=(1-x^2)(1-2ax+a^2e^2x^2),
\end{equation}
where $a=\alpha M$, $e=Q/M$, and $m=m_0P(-1)$. For $m_0\ne0$, we factor
$\chi=(1+x)^{\rho_-}(1-x)^{\rho_+}y(x)$, with
$\rho_-=|m|/[2P(-1)]$ and $\rho_+=|m|/[2P(1)]$, and expand $y$ in Jacobi
polynomials. If $\phi_i$ denotes a basis function including these endpoint
factors, the symmetric Galerkin matrices are
\begin{equation}
 K_{ij}=\int_{-1}^1\left[H\phi_i'\phi_j'
 +\left(\frac{m^2}{H}-\frac{H''}{6}\right)\phi_i\phi_j\right]dx,
 \qquad S_{ij}=\int_{-1}^1\phi_i\phi_j\,dx.
\end{equation}
The integrals use Gauss--Jacobi quadrature with exponents
$(2\rho_+-1,2\rho_--1)$ after algebraic cancellation of endpoint factors.
For $m_0=0$, we use Legendre polynomials directly. Tables~\ref{tab:generic_cmetric} and \ref{tab:extreme_cmetric} use degree $44$ and $160$ quadrature nodes in double precision, with refinement through $(d,N_{\rm quad})=(24,80),(36,120),(44,160)$. Here $d$ is the polynomial degree and $N_{\rm quad}$ the number of quadrature nodes. The largest change is $3.74\times10^{-14}$. Tables~\ref{tab:near_extreme_cmetric} and \ref{tab:extended_extreme_cmetric} add $(d,N_{\rm quad})=(64,240)$ and use this last value; their largest refinement change is $2.28\times10^{-13}$. Differences below $10^{-12}$ are reported at the numerical comparison floor. Spherical and axisymmetric exact-limit controls have absolute differences below $4.3\times10^{-13}$.

\paragraph{Instanton coefficients and truncation.}
The formal logarithm of the Young-diagram sum is formed coefficient by
coefficient before taking the NS limit; we do not take the logarithm of an
evaluated finite partition polynomial. For the generic spectra we use
$\varepsilon=10^{-26}$ with $350$ significant digits and repeat at
$10^{-30}$ with $430$ digits. At $m_0=0$, the resonant momentum
$b=\ell+1/2$ is approached from generic $b$, with shifts $10^{-16}$ and
$10^{-18}$ in these two checks. For the extreme spectra, the regulator and
precision pairs are $(10^{-22},200)$ and $(10^{-28},260)$.
These two tables have a maximum regulator/precision change below $7\times10^{-16}$. For Table~\ref{tab:near_extreme_cmetric}, including its $Q=M$ row, the pairs are $(10^{-34},650)$ and $(10^{-40},800)$, with axisymmetric momentum shifts $10^{-18}$ and $10^{-20}$. For Table~\ref{tab:extended_extreme_cmetric}, they are $(10^{-28},350)$ and $(10^{-34},430)$. The largest change in these additional spectra is $4.51\times10^{-18}$. Truncation errors are measured independently against the angular benchmarks. The quiver ordinary-point reduction was also checked
against independently assembled two-node coefficients through total degree
six. Exact rational checks at generic masses verify the same cancellation and the matching reduced coefficients through $\kappa^3$. Equation~\eqref{eq:regular_prepotential_reduction} is checked through total degree six at generic masses, including vanishing endpoint masses. The coordinate and accessory relations in \eqref{eq:physical_collision_frame}--\eqref{eq:physical_collision_limits} are also verified symbolically.

For the generic connection tests, the truncated Matone equations are solved
directly for $(b_1,b_2)$ on the branch continuous from the zero-instanton
solution. The mass derivatives in the connection formula are evaluated
before this substitution, holding $(b_1,b_2)$ fixed. Smaller NS regulators
and higher precision leave the displayed digits stable. The spectral tables fix the quantized momentum before truncation; the connection tables solve the truncated Matone system directly.

\paragraph{Independent connection coefficients.}
Write $s=z/t$. At each finite endpoint we expand a solution with unit
Frobenius leading coefficient and match the two bases at $s=1/2$ using
Wronskians. If $Q(s)=\sum_{j=-2}^{\infty}q_j s^j$ and
$\psi=s^r\sum_{n\ge0}c_ns^n$, the recurrence is
\begin{equation}
 c_0=1,\qquad
 c_n=-\frac{\displaystyle\sum_{j=-1}^{n-2}q_jc_{n-j-2}}
 {(r+n)(r+n-1)+q_{-2}}.
\end{equation}
For the positive-$t$ tests, the branch is
$(z-t)^r=e^{i\pi r}(t-z)^r$ on $0<z<t$. The original local normalization
is restored by the factors $t^r$. Increasing the series from $180$ to $240$
terms, precision from $80$ to $100$ decimal digits, and changing the matching
point from $s=1/2$ to $0.45$ changes the large coefficient by less than
$8\times10^{-48}$ for all the tabulated targets. The finite-$\eta$
Wronskian differences approach the confluent result as $O(\eta^{-1})$.

For the additional physical spectra, we also match the regular Frobenius solution at each endpoint. The reduced angular coordinate is $s=w/T$ in the non-extremal Heun frame \eqref{eq:heun_reduction}, and $s=z/t$ in the extremal frame \eqref{eq:reduced_che}. Writing $v_i=(\psi_{i,\mathrm{reg}},\partial_s\psi_{i,\mathrm{reg}})$ at $s=1/2$, we use $W_{\mathrm{norm}}=W(\psi_{0,\mathrm{reg}},\psi_{1,\mathrm{reg}})/(\|v_0\|_2\|v_1\|_2)$. This normalized Wronskian is evaluated at the Galerkin eigenvalue with $200$ terms at $80$ digits and $260$ terms at $100$ digits. Its largest absolute value is $1.06\times10^{-13}$; moving $\lambda$ by $\pm10^{-5}$ gives opposite Wronskian signs for every checked mode. The term/precision refinement changes this normalized value by less than $1.5\times10^{-61}$. The calculation files include these controls for both $m_0=0$ and $m_0\ne0$, and verify the symmetry $\lambda(m_0)=\lambda(-m_0)$ at fixed $\ell$.

The accompanying calculation files contain the Young-diagram implementation,
angular Galerkin solver, Frobenius recurrences, geometric and sewing checks,
full order-dependent data, and commands to reproduce the tables.

\bibliographystyle{JHEP}
\bibliography{YWZ}

\end{document}